\input psfig
\documentstyle[twoside,fleqn,espcrc2]{article}

\newcommand{\AmS}{{\protect\the\textfont2
  A\kern-.1667em\lower.5ex\hbox{M}\kern-.125emS}}

\title{Heavy-Light Wavefunctions in Lattice QCD
       \thanks{Talk presented by H.~Thacker}}
\author{A. Duncan\address{Department of Physics and Astronomy,
        University of Pittsburgh, Pittsburgh, PA 15620}
        E. Eichten
        \address{Fermi National Accelerator Laboratory,
        P.O. Box 500, Batavia, IL 60510},
        and
        H. Thacker
        \address{Department of Physics,
        University of Virginia, Charlottesville, VA 22901}}

\begin{document}

\begin{abstract}
Using a multistate smearing method, Coulomb gauge wave functions of heavy-light
mesons are studied in lattice QCD. Wave functions
for the ground state,  the first
radially excited S-wave state, and the lowest P-wave states of a
heavy-light meson are calculated in quenched approximation.
The results are found to be in remarkably good agreement with
the predictions of a simple relativistic quark model.
\end{abstract}

\maketitle

\section{Introduction}

The evolution of lattice gauge theory techniques has greatly enhanced our
understanding of quark-gluon dynamics in QCD.
Heavy-light mesons provide an ideal laboratory
for lattice QCD studies.
The static approximation ($m_Q \rightarrow \infty$) in which the heavy
quark propagator is replaced by a straight time-like Wilson line
provides a framework which allows a
quantitative study of masses, decay constants, mixing amplitudes,
and electroweak form factors.[1]
Since heavy-light mesons have only one dynamical
light (valence) quark, these systems are also well suited to the
study of constituent quark ideas [2] and the chiral
quark model [3].

In view of the success of the nonrelativistic (NR) potential
model for heavy $Q\bar Q$ mesons,
one interesting question for heavy-light systems is the
nature and extent of the deviation from the NR potential
picture as one of the quarks becomes light.
Here we present results of a numerical lattice study of this question.
Our findings support a surprisingly simple answer.
The Coulomb gauge wave functions obtained in lattice QCD agree,
within the accuracy of our calculations, with the results of a simple
relativistic generalization of the NR quarkonium potential model.
It is only necessary to replace the NR kinetic energy term in the
Hamiltonian by its relativistic form, leaving the NR
potential unchanged.  The
only adjustable parameter is the quark
mass parameter $\mu$.
This description holds down to fairly small values of the current
quark mass, corresponding to a pion mass of approximately $300 MeV/c^2$, well
into
the region where the NR description fails.

\section{Wavefunctions in Lattice QCD}

In lattice QCD, the properties of hadronic states are studied
using correlation functions of operators which couple to the state.
Originally
local operators were used. More recently smeared (non-local) operators have
been found to improve the ability to extract the masses of meson and baryon
ground states [4].  Many of the present studies have been
done with configurations and propagators
fixed in Coulomb gauge and operators which smear the position
of the quark field uniformly over a spatial cube of variable size.
However a constant cube of any size is a very crude approximation to the
ground state wave function [5].
Hence, the propagator generally has significant
contamination from higher states
out to times large compared to the inverse of the energy splitting between the
ground state and the lowest excited state.

This is a particular problem in the study of
heavy-light correlators because they become noisy
rather rapidly in time.
Unfortunately, this is an unavoidable feature of heavy-light
systems [6,7].
Recently a multistate smearing technique
has been proposed [6]
which allows the extraction of
the properties of heavy-light states
from relatively short times.

The details of the multistate smearing method have been
presented elsewhere[6]. By choosing an appropriate orthonormal set of smearing
functions and diagonalizing the corresponding matrix of correlators, one
obtains the wave functions of not only the lowest lying state in a given
channel,
but also of radially excited states. Here we define the wave function to be
the vacuum-to-one-particle matrix element,
\begin{equation}
\Psi(\vec{r}) = \sum_a\langle 0|q_a(\vec{r},0) Q_a^{\dag}(0,0)|B\rangle
\end{equation}
where $|B\rangle$ is the state of interest. The sum is over color, and
spin labels are supressed.

Here we discuss the wave functions obtained for the 1S and 2S
levels of the S-wave pseudoscalar meson as well as a preliminary study of
the 1P state.[8] The main emphasis will be on the remarkable quantitative
agreement between the lattice QCD wave functions and those obtained from
a simple relativistic quark model Hamiltonian. The results for heavy-light
decay constants and spectroscopy will be presented at this conference
by Eichten.[9]

The investigation used an existing set of 50 configurations
(each separated by 2000 sweeps) generated by ACPMAPS on
a $16^3\times 32$ lattice at $\beta = 5.9$. The configurations
were fixed to Coulomb gauge and light quark propagators
with $\kappa = .158$ were used.
Only the four lowest energy smearing
functions were included ($N = 4$).

\section{Relativistic Quark Model}

The optimized wave functions obtained from our lattice data by the multistate
smearing method turn out to be, within errors, the same as the eigenfunctions
of a lattice version of the spinless, relativistic quark model Hamiltonian,
which
we will now define. In the absence of gauge fields, the free quark Hamiltonian
can be exactly diagonalized by introducing momentum space creation and
annihilation
operators for quarks and antiquarks. In the continuum,
\begin{eqnarray}
H_0 = \int\frac{d^3p}{(2\pi)^3} \sqrt{\vec{p}^2+\mu^2} \sum_i[
\alpha^{\dag}_i(p)\alpha_i(p)\nonumber\\
 + \beta^{\dag}_i(p)\beta_i(p)]\label{eq:H0}
\end{eqnarray}
where the sum is over spin and color labels. In terms of the covariant quark
propagator, the particle and antiparticle operators are associated with
propagation
forward and backward in time, respectively. Since $H_0$ contains no pair
creation
($\alpha^{\dag}\beta^{\dag}$) terms, it is possible to formulate the eigenvalue
problem as that of a one-body operator,
$H_0 \rightarrow \sqrt{\mu^2-\nabla^2} $
If we now turn on the gauge interaction and introduce a heavy-quark, static
color
source, the description of the bound light quark becomes, in principle,
drastically
more complicated. We know that, in the limit $\mu\gg\Lambda_{QCD}$ where the
dynamical
quark becomes heavy, the primary effect of the color source is to introduce a
static,
confining potential $V(r)$ whose form is well-measured and consistently given
by both
$Q\bar{Q}$ phenomenology and lattice QCD,
\begin{equation}
H_0 \rightarrow H = H_0 + V(r) \label{eq:SRQM}
\end{equation}
At this stage, the Hamiltonian can still be regarded as a one-body
operator[10]. As the mass of the quark becomes light, one expects more
complicated effects
arising from the gauge interaction which render the Hamiltonian eigenvalue
problem
intractable. These effects include the creation of gluons and light $q\bar{q}$
pairs,
as well as the exchange of transverse and non-instantaneous gluons with the
static
source. From the numerical results presented in the next section, we conclude
that
these effects are relatively small, and that the heavy-light meson system is
well-described
by the Hamiltonian (6), which we will refer to as the spinless relativistic
quark
model (SRQM).

\begin{figure}[htb]
\psfig{figure=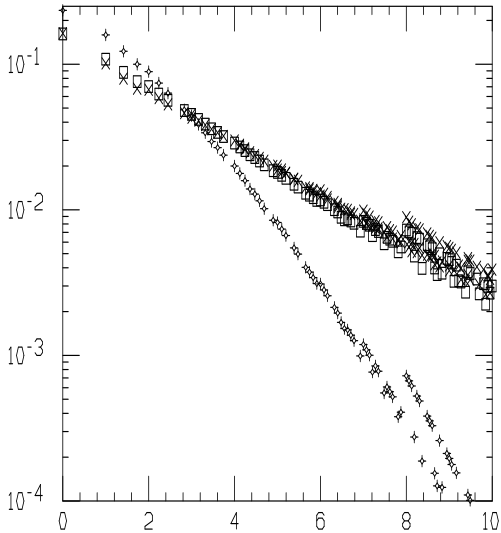,height=3.0in,width=2.8in}
\caption{Comparison of the 1S state in LQCD ($\times$'s) with the NRQM
(+'s) and the SRQM (boxes).}
\label{fig:toosmall}
\end{figure}

The construction of explicit eigenfunctions of the SRQM Hamiltonian is easily
accomplished by a numerical procedure. First the operator $H$ is discretized on
a 3D
lattice by replacing the spatial derivatives with finite differences
The potential energy V(r) is just the static energy
measured on the same configurations used to study
the heavy-light spectrum.
Then the resolvent
operator $(E-H)^{-1}$ acting on a source vector $\chi$ is computed by a
numerical
matrix inversion (conjugate gradient) algorithm. Finally, the parameter $E$ is
varied
to find the poles in the output vector $(E-H)^{-1}\chi$. The location of the
pole is
an eigenvalue of $H$, and its residue is the corresponding eigenfunction. In
the next
section we compare the wave functions obtained in this way from the SRQM
Hamiltonian
with the lattice QCD results.

\section{Comparison of Wavefunctions}

\begin{figure}[htb]
\psfig{figure=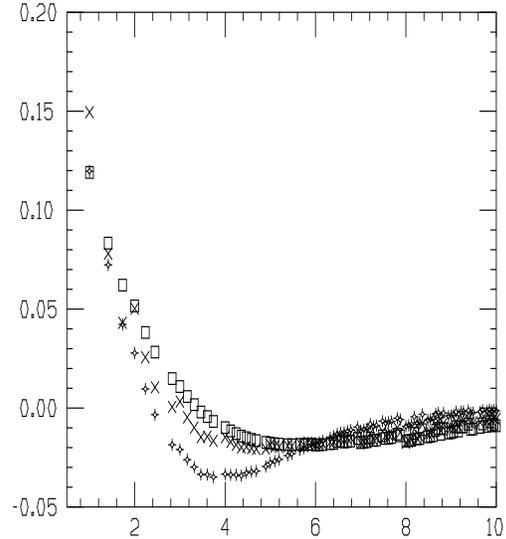,height=3.0in,width=2.8in}
\caption{Comparison of the 2S state in LQCD ($\times$'s) with the NRQM
(+'s) and the SRQM (boxes).}
\label{fig:toosmall}
\end{figure}

\begin{figure}[htb]
\psfig{figure=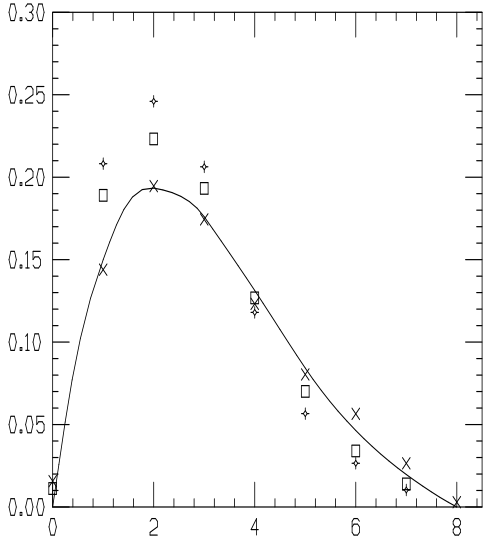,height=3.0in,width=2.8in}
\caption{The 1P state in LQCD extracted from T=2 (+'s), T=4 (boxes) and,
T=6 ($\times$'s}).
\label{fig:toosmall}
\end{figure}


Using the four state smeared correlator described in
section 2 an initial study for
the S-wave
channel was carried out.
After some iterative improvement of the smearing functions, it was found that
the value $\mu = .23$ for the dimensionless mass parameter in the SRQM
Hamiltonian gave the best agreement with the lattice QCD wave functions with
$\beta = 5.9, \kappa = .158$. In Fig.~1 the LQCD wave function
is plotted with the SRQM wave function. For comparison, the
nonrelativistic (NR) Schrodinger wave function (obtained by replacing the
relativistic kinetic term by $p^2/2m$) is also plotted. The mass
parameter in the NR Hamiltonian was adjusted to give the same slope at
the origin in the ground state wave function. Notice that, for large r,
the QCD and SRQM wave functions both fall exponentially. On the other
hand, the NR wave function falls faster
than exponentially ($exp(-\alpha r^{\frac{3}{2}}$), as expected from the
behavior of the analytic solution in a pure linear potential (Airy function).
Remarkably, by including the relativistic kinetic term, the SRQM wave
functions are brought into excellent agreement with those of lattice QCD,
without changing the potential from its nonrelativistic form.

In Fig.~2 we plot the excited 2S state from LQCD along with the corresponding
wave functions from the SRQM and the NR model.
The QCD wave function is somewhat more peaked at the origin, however,
the overall agreement between QCD and the SRQM is excellent.
Here, there are no adjustable parameters, $m$ being already fixed from
the 1S state fit.
Finally, in Fig.~3 we show some preliminary results of a study
of the 1P state. Here the solid line is the 1P wave function from the SRQM.
The data points depict the evolution of the P-wave LQCD radial wavefunction
extracted from time slices $T=2$ (+'s), $T=4$ (boxes), and
$T=6$ ($\times$'s), starting with an approximate guess for the initial smearing
function. The ansatz for the initial smearing function used here was a simple
$re^{-\alpha r}$ form. As the LQCD wave function evolves in Euclidean time,
it appears to approach a true eigenstate whose wavefunction again agrees
remarkably well with the SRQM result, with no adjustable parameters.

\section{Discussion}

Additional studies are in
progress using a variety of lattice sizes, gauge coupling strengths, and light
quark masses.  Preliminary results of these studies are fully consistent with
the conclusions presented here.
The agreement of lattice QCD with the SRQM wave functions suggests
that the relativistic propagation of the light valence quark is
the most important effect which must be included in a description of
heavy-light mesons.
Other field theoretic effects such as the presence
of multibody components of the wavefunction (containing gluons along
with light $q\bar{q}$ pairs arising, in quenched approximation, from the
propagation of the valence quark backward in time) are of less
quantitative importance in determining the shape of
the valence quark wave function.
Further numerical studies of the connection between lattice QCD
and the relativistic quark model are planned.

\section{Acknowledgements}

 We thank George Hockney, Aida El Khadra, Andreas Kronfeld, and
Paul Mackenzie for joint lattice efforts without which this analysis
would not have been possible.  We also thank Chris Quigg for
useful discussions. The numerical calculations were performed on the
Fermilab ACPMAPS computer system developed by the CR\&D department in
collaboration with the theory group. This work was supported in part by
the U.~S.~Department of Energy under grant No. DE-AS05-89ER40518.

\end{document}